\documentclass[12pt]{article}
\usepackage{amssymb,amsmath,latexsym,graphics,authblk,url,setspace,caption}
\usepackage[pdftex]{graphicx}
\usepackage{epstopdf} 
\usepackage{enumerate}
\usepackage{breqn}

\usepackage[square,sort,comma,numbers]{natbib}

\providecommand{\keywords}[1]{\textbf{\textit{Keywords---}} #1}

\usepackage[letterpaper, portrait, margin=1.5in]{geometry}

\begin{document}
\pagestyle{empty}
\title{Practical data synthesis for large samples}
\author{Gillian M. Raab\footnote{gillian.raab@ed.ac.uk}, Beata Nowok\footnote{beata.nowok@ed.ac.uk} and Chris Dibben\footnote{chris.dibben@ed.ac.uk}}
\affil{School of Geosciences, University of Edinburgh, Edinburgh, UK}
\date{\today}

\maketitle
\abstract{We describe results on the creation and use of synthetic data that were derived in the context of a project to make synthetic extracts available for users of the  UK Longitudinal Studies.
	A critical review of existing methods of inference from large synthetic data sets is presented. We introduce new variance estimates for use with large samples of completely synthesised data that do not require them to be generated from the posterior predictive distribution derived from the observed data and can be used with a single synthetic data set. We make recommendations on how to synthesise data  based on these findings.  An example of synthesising data from the Scottish Longitudinal Study is included to illustrate our results.}\\

\setlength{\parindent}{3em}
\setlength{\parskip}{0.3em}


\section[Intro]{Introduction and background}
\subsection{Synthetic data for disclosure control}
National statistics agencies and other groups collect large amounts of information about individuals and organisations. Such data can be used to understand population processes so as to inform policy and planning. 
The cost of such data can be considerable, both for the collectors and the subjects who provide their data. 
Because of confidentiality constraints and guarantees to data subjects full access to such data is often restricted to the staff of the collection agencies. 
Traditionally, data collectors have used anonymization along with simple perturbation methods such as aggregation, top-coding, record-swapping, suppression of sensitive cells or adding random noise, to prevent the identification of data subjects. 
Advances in computer technology and search techniques have illustrated how such measures may not prevent disclosure \cite{Ohm_2010}. 
Additionally, examples show that these ad hoc disclosure procedures may compromise the conclusions that can be drawn from such data \cite{Winkler_2007,elliot_purdam_2007}.
   
Synthetic data, which retains the essential features of the actual data with some or all of the values replaced by simulations from probability distributions, can be a way of overcoming these difficulties. There is now an extensive literature on this topic \cite{RRR_2003,Reiter_2002,Reiter_RSS,Caiola_and_R_2010,DR_jasa_2010,reiter_cart,kinn_R2010,Drechsler_2012}, much of it summarised in the monograph by Drechsler \cite{Drechsler_book}, as well as several  initiatives, e.g.  \cite{synLBD,SIPP}, which make synthesised data available to researchers.
 
 \subsection[LSs]{Application to the UK Longitudinal Studies}
 The England and Wales Longitudinal Study (ONS LS) \cite{hat} the Scottish Longitudinal Study (SLS) \cite{boyle} and the Northern Ireland Longitudinal Study (NILS) \cite{or} are rich micro-data sets linking samples from the national census in each country to administrative data (births, deaths, marriages, cancer registrations and other sources) for individuals and families across several decades. Researcher access to the UK Longitudinal Studies  (LSs) is highly restricted due to confidentiality and legal constraints. Thus  the three LSs have a small number of users compared to other census data products. Synthetic data with no real individuals, but which mimic the real data and preserve the relationships between variables and transitions of individuals over time, could be made available to accredited researchers to analyse on their own computers.
 
 Although the LSs are pseudo-random samples from the censuses, with linked administrative data, people use the data to draw general conclusions about social processes, rather than to infer anything about the census totals. Thus inferences from the LSs generally assume that the data are samples from a hyper-population model that might have generated the data. Methods appropriate for simple random sampling (SRS) are generally used, except for analyses that involve clustering, e.g. in families or by geographic area, when hierarchical models may be used. 

 Other projects that make synthetic data available have been based around a single data set, so that synthesised data sets are created once and extracts from them are supplied to users. Every user of the LSs has a customised, linked data set made available to them. Thus a new synthesis is required for every researcher. To make it possible for LS support staff to provide synthesised extracts for researchers we have developed  the \textbf{R}  package $synthpop$   \cite{synthpop_jss} which is now available free to any user.\footnote{Its initial development was part of the SYLLS (Synthetic Data Estimation for UK Longitudinal Studies) project  \url{http://www.lscs.ac.uk/projects/synthetic-data-estimation-for-uk-longitudinal-studies/}  funded by the UK  Economic and Social Research Council.} As we implemented  different proposed methods of inference from synthetic data, we were led to
re-evaluate the conditions under which each is valid.  We have also derived  new variance estimates which can be used under certain conditions and which can be calculated from a single synthetic data set. 
3 
  \subsection[review]{Review and critique of methodology for inference from synthetic data}
Rubin's original proposal for synthetic data \cite{Rubin1993} saw it as an
 example of multiple imputation (MI). The first paper to provide detailed methodology for making inferences from synthetic data \cite{RRR_2003} followed this approach. 
Synthesis  was assumed to proceed by imputing all the unobserved members of a population and then selecting a synthetic sample from each synthetic population.
Because imputation is involved, it was considered mandatory to generate the synthetic populations from their posterior predictive distribution (PPD), given the observed data. 
   The variance of estimates  from synthetic data differ from those for MI by replacing the addition of the term for within-imputation variance  by its subtraction. 
   The authors (\cite{RRR_2003}, Section 2.2) attribute this difference to the sampling for each synthetic sample from its population and this explanation is expanded in \cite{RR_JASA_2007}, Section 3.1. This observation is correct in that the between-synthesis variance includes 
   a contribution from sampling from the synthesis distribution, but it is not the underlying reason for the negative sign in the variance estimate.  
   
   We will show that results for inference from synthetic data can be derived without considering it as imputation at all. 
   The negative sign arises from the need to get the contribution to the variance of the synthetic estimates from the uncertainty about the population parameters from the observed data. 
   When the estimate would be calculated in a different manner from the observed and synthetic data, we need multiple syntheses sampled from the PPD of $Y_{obs}$ so that this uncertainty contributes to the between synthesis variation.  The variance contribution from the observed data can then be obtained by subtracting the within-synthesis variance from the between-synthesis variance. 
   For the case when the same estimation methods would be used for the observed and synthetic data, the contribution from the fit to the observed data can be estimated from the within-synthesis variation. 
   This leads to new variance estimates that can be calculated from a single synthetic data set without sampling from the PPD.
 
 We assume that the observed data consist of $(Y_{obs},X)$ and all or part of $Y_{obs}$ are to be replaced by synthesised data $D_{syn}$ while the original values of $X$ are all retained and the dependence of $Y_{obs}$ on $X$ is respected in $D_{syn}$.  Those left unchanged will often include design variables such as stratum indicators or 
 weights that define the relation between the distribution of  the observed data and the population from which it has been sampled. When inference is carried out for a statistic $Q$, conditional on $X$, from $(D_{syn}|X)$  then the variance estimator derived  in \cite{RRR_2003} and our new variance estimators require that all of the observed data $Y_{obs}$ are replaced by synthetic values. Reiter \cite{Reiter_partial} derived results for making inferences when only part of $Y_{obs}$ is synthesised and Reiter and Kinney  \cite{reiter_kinney_2012} pointed out that sampling from the PPD was not needed for what they term ``partially synthetic" data. Their examples include cases where all of $Y_{obs}$ are synthesised. They distinguish fully synthetic data, which requires sampling from the PPD, from partially synthetic data, which does not, by the fact that no new units are introduced in partially synthetic data so that synthetic populations sampled from the PPD are not needed.
 
 We demonstrate that data synthesis, as usually carried out in the literature cited here, does not require the creation of synthetic populations, except in the sense of defining them in terms of some appropriate model. This is true whether or not the synthetic data includes new units that were not part of the observed data.  Thus Reiter and Kinney's distinction between fully and partially synthetic data is not meaningful.  The real reason why PPD is not required for ``partially synthetic" data is that the same estimators are used to analyse the observed  and synthetic data. This will always be true when some observed data are retained and it is also the case for the examples of ``partial synthesis" cited in the literature.  To avoid confusion with the previous terminology we will use the term ``completely synthesised" when all of $Y_{obs}$ from $(Y_{obs}|X)$ (as defined above) is replaced by synthetic data and ``incompletely synthesised" when part of the original $Y_{obs}$ is left unchanged. Although PPD is not required for incompletely synthesised data, multiple syntheses are still needed because the variance calculated from the synthetic data, as if it were real, will no longer include all of the contribution from the fit to the observed data.

In most implementations of synthetic data generation, including $synthpop$, the joint distribution is defined and synthesised in terms of a series of conditional distributions. 
This approach has been used extensively in implementations of MI, e.g. \cite{mice_jss_2011,Rag_imp,royston2004}. In MI missing values are given starting values, followed by a chain of updates when missing values for each variable are replaced by samples from their PPD given other variables.  This has led to concerns about validity when the conditionals that define each variable are not compatible \cite{vanBuuren_2007}. 
This concern does not apply to synthetic data because the joint distribution is defined by starting with one marginal distribution
and building up the joint distribution by modelling each additional variable conditional on those already synthesised. This ensures that joint distribution must exist. 
In our experience synthesising data from conditional distributions works well and allows great flexibility to produce plausible synthetic data that mimics the structure of real survey or administrative data. A more important requirement underlies all synthetic data estimation. This is the assumption that the observed data were generated from the same model that was used for synthesis. We will refer to this as the  Synthesising Distribution Assumption, or the SDA for short. The results derived in this paper only apply for large samples, making the same asymptotic assumptions as in the other literature cited, and where the  SDA  is met.

 \subsection[papsum]{Structure of this paper}
 
The structure of this paper is as follows. Section \ref{sec:inference} presents our results on inference from synthetic data. This is first developed for the case when methods appropriate for SRS can be used to obtain estimates for both the observed and synthetic data. Completely synthesised data are considered first, our new variance estimators are introduced and their properties confirmed by a simulation. Results for incompletely synthesised data are presented in subsection \ref{sec:inference:incomplete} and the extension to methods for complex samples in subsection \ref{sec:inference:complex}. 
We conclude this section by summarising our recommendations for practice based on these results. 
Section \ref{sec:practical} reports on our practical experiences of creating synthetic data using $synthpop$ and Section \ref{sec:disclosure} briefly reviews aspects of the disclosure risk and utility of synthetic data. 
An example of synthesising data from the SLS is presented in Section \ref{sec:example} and compares the different variance estimators discussed in Section \ref{sec:inference}. The final section presents a summary and proposes future directions for our work.

\section[inference]{Inference from synthetic data}
\label{sec:inference}

\subsection[complete]{Completely synthesised data}
\label{sec:inference:complete}
 For inferential purposes data are considered completely synthesised if either all of the observed data are replaced by synthetic values, or if only certain  variables, denoted by X, are left unchanged and both synthesis and inference is carried out conditional on X.
Formally, for complete synthesisis, all of $Y_{obs}$ is replaced by $D_{syn}$ which are generated as samples from the PPD of $(y|Y_{obs},X)$, where $X$ are variables that are left unchanged. If $X$ is not empty then all inference from the synthetic data must be conditional on $X$.
  
In \cite{RRR_2003} synthetic data are described as MI by first creating synthetic populations generated from the PPD of $(y|Y_{obs},X)$.  Each synthetic data set $Y^{(l)},l=1,\dots,M$ is then  generated by SRS from the $l^{th}$ population, although, as \cite{RR_JASA_2007} have pointed out, it is not necessary to generate the synthetic populations but only values for the synthetic samples. This implies that the synthesis does not make use of any detailed individual information that may be available for the non-sampled members of the  population from which $Y_{obs}$ is a sample. This has been the case, with a few exceptions e.g.\cite{samp_w_syn},  for most practical applications of synthetic data. 

In \cite{RRR_2003} the target for inference, $Q$, is defined as a function of the finite population quantities $(X,Y)$. However, these authors comment in Section 2.1 that, when population values of $X$ are not available or cannot be released, synthetic populations ``can be generated based on the posterior predictive distribution of \lq super\rq  or \lq future\rq  populations". 
This suggests defining $Q$ as a function of the parameters, $\theta$, of some parametric distribution assumed to have generated the observed data so that $Y_{obs} \sim  f(y|X,\theta)$ for some distribution $f$. Based on this approach, we provide a derivation of the variance estimator ($T_M$) derived in \cite{RRR_2003} without the requirement to create synthetic populations, except in the sense of defining their distributions, and without considering synthesis as an example of MI.
 
 We assume that the synthetic data, $D_{syn}$, to replace  $Y_{obs}$, are produced by fitting the parameters of the distribution $f(y|X,\theta)$ to the observed data $(X_{obs},Y_{obs})$ and generating $D_{syn}$, ($Y^{(1)},Y^{(2)},...,Y^{(l)},...,Y^{(M)}$), as $M$ simple random samples from the PPD of $(y|Y_{obs},X,\theta)$. The analyst wishes to make inferences about a quantity $Q$, conditional on $X$, given only the synthetic data. The $X$ variables forming part of each synthetic data set may consist of  $X_{obs}$, either unchanged 
  or altered in some way, as has been discussed in \cite{mitra_weighting} in relation to survey weights.
 From each $(Y^{(l)},X)$ we calculate $q^{(l)}$ and $v^{(l)}$, the estimate of $Q$ and an estimate of its variance, conditional on $X$, as if  $(Y^{(l)},X)$ were the observed data generated as a simple random sample. We aim to make inferences for $Q$ based on  
\begin{center}
$\bar{q}_{M} = \sum_{l=1}^{M}{q}^{(l)}/M$,~~~$\bar{v}_{M} = \sum_{l=1}^{M}{v}^{(l)}/M$~~~and
\end{center}
\begin{center}
$b_M = \sum_{l=1}^{M}{({q}^{(l)}-\bar{q}_{M})}^2/(M-1)$.\\
\end{center}
\noindent
The $l^{th}$ synthetic data set, $Y^{(l)}$ can be considered as a simple random sample from  $f(y|\hat{\theta}^{(l)},X)$ where $\hat{\theta}^{(l)}$ is a sample from the posterior of     $(\theta|Y_{obs},X)$. Since we can write $Q=Q(\theta)$ the population quantity corresponding to $\hat{\theta}^{(l)}$ can be written as $\hat{Q}^{(l)} = Q(\hat{\theta}^{(l)})$. This is the quantity that will be estimated by $q^{(l)}$ from the $l^{th}$ synthetic data set. Thus $\hat{Q}^{(l)}$ takes on the role of the estimate from the $l^{th}$ population in \cite{RRR_2003}. 
 We make the following  assumptions which are equivalent to those in \cite{RRR_2003}, but are stated more explicitly.

\begin{enumerate}
  \item {Diffuse priors are assumed for all parameters that are non-informative over the range where the posterior has any support from the data.} 
   \item{The estimate $q^{(l)}$ from $Y^{(l)}$ is unbiased for $\hat{Q}^{(l)}$ and asymptotically Normal with respect to repeated sampling from the $l^{th}$ population, $f(y|\hat{\theta}^{(l)},X)$, with variance estimate ${v}^{(l)}$  unbiased for $V^{(l)}$ and the sampling variance of ${v}^{(l)}$ is negligible.}
  \item{The 
  	variation of  $V^{(l)}$ across the $M$ syntheses is negligible. Thus, using assumption 2, we have  ${v}^{(l)}  \approx{~} {V}^{(l)} \approx{~} V$  for all $l$.}
        \item{The estimates $\hat{Q}$ and $\hat{\theta}$ from $(Y_{obs},X)$ are unbiased for $Q$ and $\theta$ and asymptotically Normal. The variance of $\hat{Q}$ is $U$, estimated by $\hat{U}$ with sampling variability negligible relative to that of $\hat{Q}$, so we can write $\hat{U} \approx{~} U$.} 
\end{enumerate}

\noindent{
The estimates of $\theta$  in assumption 4. are those that would be used by the synthesiser to obtain the posterior distributions from which the  $\hat{\theta}^{(l)}$ are sampled and the synthetic populations  $f(y|\hat{\theta}^{(l)},X)$, are generated. Since $(Y_{obs},X)$, unlike $(D_{syn},X)$, may not have been generated by simple random sampling the variance $U$ may not be the same as $V$. From assumptions 1. to 3. the posterior of each $q^{(l)}$ conditional becomes
\begin{equation*}  
Pr(\hat{Q}^{(l)}|q^{(l)}) = N(\hat{Q}^{(l)},V).
\end{equation*} 
Since $Q=Q(\theta)$ and  $\hat{Q}^{(l)}=Q(\hat{\theta}^{(l)})$ where each $\hat{\theta}^{(l)}$  is a sample from the the PPD of $(\theta|Y_{obs},X)$ it follows that each $\hat{Q}^{(l)}$ a sample from the PPD of $(Q|Y_{obs},X)$.Thhus from assumption 4.,
\begin{equation*}  
(\hat{Q}^{(l)}|Y_{obs},X) \sim N(\hat{Q},U).
\end{equation*}
These conditions ensure that  $\bar{q}_{M}$  is asymptotically unbiased for $Q$ since each $q^{(l)}$ has expectation $\hat{Q}^{(l)}$ and each  $\hat{Q}^{(l)}$ has expectation $\hat{Q}$ which is itself unbiased for $Q$ from assumption 4. 
Now the required posterior distribution of $Q$, given the synthetic data becomes
\begin{equation*}
Pr(Q|D_{syn},X)= Pr(Q|\bar{q}_{M},X)
\end{equation*}
\noindent{If $\bar{Q}_{M} =\sum_{l=1}^{M}{\hat{Q}}^{(l)}/M$ we can obtain an expression for the variance of $Q$ by considering the distribution of $(Q|\bar{q}_{M},X)$ conditional on $\bar{Q}_M$}
\begin{equation}
var(Q|\bar{q}_{M},X)=var[E(Q|\bar{q}_{M},X,\bar{Q}_{M})|\bar{q}_{M},X]
+ E[var(Q|\bar{q}_{M},X,\bar{Q}_{M}) |\bar{q}_{M},X].
\label{eq:1}
\end{equation}
\noindent{ 
 The expectation in the first term is $\bar{Q}_{M}$ and since the posterior of $Q$ given $q^{(l)}$ is $N(\hat{Q}^{(l)},V)$ and $(q^{(l)}|\hat{Q}^{(l)})$ is independent of $(q^{(j)}|\hat{Q}^{(j)})$  $l\neq{j}$, the posterior of 
$Q$ given  $\bar{q}_M$ is $N(\bar{Q}_M,V/M)$. Thus \eqref{eq:1} becomes 
\begin{equation}
var(Q|\bar{q}_{M},X)=var(\bar{Q}_{M})+V/M.
\label{eq:2}
\end{equation}
Considering the distribution of $\bar{Q}_{M}$ conditional on $\hat{Q}$ we get 
\begin{equation}
var(Q|\bar{q}_{M},X)=
var[E(\bar{Q}_{M}|\hat{Q})]+
E[var(\bar{Q}_{M}|\hat{Q})]+
V/M.
\label{eq:3}
\end{equation}
The expectation in the first term is $\hat{Q}$ with variance $U$ from assumption 2.  
Since the variance of $(Q^{(l)}| \hat{Q})$ is $V$ and  $(Q^{(l)}| \hat{Q})$ and $(Q^{(j)}|\hat{Q})$ are independent $l\neq{j}$, 
$var(\bar{Q}_{M}|\hat{Q})$ is $U/M$ and \eqref{eq:3} becomes

\begin{equation}
var(Q|\bar{q}_M,X) = U+U/M+ V/M=U+(U+V)/M. \label{eq:4}
\end{equation}
Synthetic data can also be created with the approximate Bayesian bootstrap \cite{ABB} and in this case these results follows immediately  because estimating $Q$ from 
the first bootstrap stage will generate $\hat{Q}^{(l)}\sim N(\hat{Q},U)$ and estimating from the second bootstrap sample will give $q^{(l)}  \sim N(\hat{Q}^{(l)},V)$. We can also understand this result by expressing the devation of $Q$ from its estimate given $D_{syn}$ as the sum of three independent deviations with variances given by the terms in equation \eqref{eq:4}.
\begin{equation}
(Q-\bar{q}_M) = (Q-\hat{Q})+(\hat{Q}-\bar{Q}_M)+(\bar{Q}_M-\bar{q}_M) \label{eq:5}
\end{equation}
\noindent
In order to make use of \eqref{eq:4} we need to replace $U$ and $V$ by unbiased estimates. We have $\bar{v}_{M}$ for $V$ from assumption 4, but for $U$ we need to consider the expectation of $b_M$,
\begin{dmath}
E(b_M)=  E\{\sum_{l=1}^{M}{[({q}^{(l)}-Q^{(l)})-(\bar{q}_M-\bar{Q}_M)]}^2/(M-1)\}+ E[\sum_{l=1}^{M}{(Q^{(l)}-\bar{Q}_{M})}^2/(M-1)]=V+U    
\end{dmath}
\noindent
so that $b_M-\bar{v}_M$ has expectation $U$ and  substituting these expectations into \eqref{eq:4} we get the asymptotically unbiased variance estimator
\begin{equation*}
T_M = b_M(1+1/M)-\bar{v}_M.   
\end{equation*} 
These results hold for large samples but, as Little \cite{Little2015} has pointed out, asymptotics may fail for some inferences even for very big data sets. We also require the assumption that the population size is large enough for the finite population correction factor (fpc) to be ignored. This condition was not mentioned explicitly in \cite{RRR_2003}, but their methods require it. Note that this is true even when whole synthetic populations are generated. The fpc will not reduce the variance of samples from synthetic populations because the totals of elements contributing to estimators will not be constant across populations and will not usually correspond to the totals for the population from which the observed data were sampled.

In the derivation of variance estimators for MI (\cite{rubin_book_1987}, Chapter 3) it is important that the between-imputation variance uses the PPD of the missing observations, given those observed, so that all sources of uncertainty about the missing observations are included. This is not the case for $Y_{obs}$ which are not missing, but only unavailable to the analyst, and contribute to the posterior of $(Q|\bar{q}_M)$. Thus  there is no a-priori reason to sample $Y^{(l)}$ from the PPD of $Y_{obs}$. However, we need  multiple samples from the PPD of $Y_{obs}$ to obtain an expression with expectation $U$ to substitute into \eqref{eq:4}. 
Our development so far requires that the synthetic data are generated by SRS, but  the observed data can be a SRS or a complex sample with, in the latter case, appropriate methods of inference (e.g. as described in \cite{lumley_book}) 
used in the estimation from $Y_{obs}$. We will show in the next section that, if $Y_{obs}$ is a SRS, a simplifications of the variance estimator is possible which can be calculated from a single synthetic data set so that multiple samples from the PPD of $Y_{obs}$ are not required.

\subsection[new]{New simple variance estimators for completely synthesised data}\label{sec:inference:new} 
When $Q$  is estimated
from $Y_{obs}$ by methods 
appropriate for SRS and the sample sizes for the observed and synthetic data are the same, we can see from assumptions 3 
that $\hat{U}\approx{~}V$, since $\hat{V}^{(l)}\approx{~}V$, and the assumption that $V^{(l)}$ 
does not vary across synthetic data sets, implies that it will also be the same for the observed data. Thus we can replace $U$ by $V$ in \eqref{eq:4} leading to 
\begin{equation*}
var(Q|\bar{q}_M,X)=V+2V/M
\end{equation*} 
\noindent{and the simple variance estimator}
\begin{equation*}
T_{s(PPD)} = \bar{v}_M {(1+2/M)}.  
\end{equation*}
\noindent{Since we no longer need an expression with expectation $U$ multiple syntheses and sampling from the PPD of $Y_{obs}$ are no longer needed. If we replace sampling from the PPD of $Y_{obs}$ to obtain the parameters to generate the synthetic data sets with sampling from $f(y|\hat{\theta})$, then $Q^{(l)}$ will be replaced by $\hat{Q}$ for all $l$, and the central term in equation \eqref{eq:4} drops out and \eqref{eq:5} becomes
	\begin{equation*}
	(Q-\bar{q}_M) = (Q-\hat{Q})+(\hat{Q}-\bar{q}_M) 
	\end{equation*}
	 to give a simple variance estimator for synthesis without PPD
\begin{equation*}
T_{s} = \bar{v}_M (1+1/M).   
\end{equation*}
In the more general case when the sample size for the observed data is $n$ and that for the synthetic data is $k$, these expressions become
\begin{equation*}
T_{s(PPD)} = \bar{v}_M (k/n+(1+k/n)/M)   
\end{equation*}
and
\begin{equation*}
T_{s} = \bar{v}_M (k/n+1/M).   
\end{equation*}
Note that the estimators $T_{s(PPD)}$ and $T_{s}$ can be computed from a single synthetic data set ($M=1$). 

The properties of these new estimators were confirmed by simulations one of which is described in Appendix \ref{sec:appendix:srs}. The new estimators were shown to be unbiased and to give the correct coverage estimates. The precision of the new variance estimates  is many times better than that of $T_M$, which can give very unsatisfactory results for small $M$.
This simulation also evaluates the estimator $T_p$, which is discussed in Section \ref{sec:inference:incomplete}.

Klein and Sinha \cite{KleinSinha2015a,KleinSinha2015b,KleinSinha2015c} have derived exact finite-sample results when the observed data are estimated by likelihood methods assuming SRS.
 Their examples include estimation of the mean for univariate and multivariate Normal distributions and linear regression, with the dependent variable synthesised but the predictors  unchanged. 
 These would all be considered as completely synthesised, since for the linear regression inference is conditional on the unchanged $X$ variables. In each case estimates are derived both with and without sampling from the PPD, with the term ``plug-in sampling" used for the latter. They note that both estimators can be used for a single synthetic data set. For all of their examples their estimates converge for large samples to $T_s$ for plug-in sampling or to $T_{s(PPD)}$ for sampling from the PPD.

\subsection[incomplete]{Incompletely synthesised data}
\label{sec:inference:incomplete}
For inferential purposes data are considered incompletely synthesised when some observed variables, and/or subsets of variables remain in $D_{syn}$ and inference is required for $Q$ that is not conditional on the unchanged data. In this case $(Q^{(l)}-\hat{Q})$  will no longer have variance $U$, so the estimators $T_M$, $T_{s(PPD)}$ and $T_s$ will no longer be unbiased. Reiter \cite{Reiter_partial} demonstrates this and derives a new estimator $T_p$ for incompletely synthesised data, and illustrates the important point that the synthetic data must be based on models fitted only to the data that are to be replaced.  The development is similar to our Section \ref{sec:inference:complete}. To reproduce it we need to introduce an expression for $E[(Q^{(l)}-\hat{Q})^2] = W$, where we expect $W<U$, since only part of the data are replaced. For sampling from the PPD of the part of the data to be synthesised equation \eqref{eq:4} becomes 
\begin{equation}
var(Q|D_{syn}) = U + (W+V)/M  
 \label{eq:7}
\end{equation}
and the  expectation of $b_M$
\begin{equation}
E(b_M) = V+W   \label{eq:8}
\end{equation}
and, as before, $\bar{v}_M$ has expectation V. When some observed data remain in the synthetic data, its structure must be the same as that of the observed data, so the analyst with synthetic data will use the same methods as if the observed data were available. Thus we have $U \approx{~} V$, as for the new variance estimators in Section \ref{sec:inference:new}. Substituting the unbiased estimate from \eqref{eq:8} into \eqref{eq:7} we obtain 
 \begin{equation*}
 T_p = \bar{v}_M+b_M/M.   
 \end{equation*}
Reiter's derivation of $T_p$ \cite{Reiter_partial} used PPD but, subsequently, Reiter and Kinney \cite{reiter_kinney_2012} showed that PPD was not required for what they termed ``partial synthesis" and that the estimator $T_p$ could be used in this case. Their argument was based on inference for a Normal mean from completely synthesised data, as well as results from a simulation with incompletely synthesised data.

That the estimator $T_p$ is also valid without PPD can be seen by noting that dropping the PPD step removes $V$ from equations \eqref{eq:7} and \eqref{eq:8} and thus leaves  $T_p$  unchanged as an unbiased estimator of $var(Q|D_{syn})$.
Reiter and Kinney argue that PPD is not needed here because synthetic populations are not generated for their examples, and only the observed units are retained in the synthetic data. They term this ``partial synthesis" even when all of $Y_{obs}$ are synthesised. We have shown here that model-based synthesis never requires the generation of synthetic populations and the criterion for when PPD is needed is that different methods of inference are used for the observed and synthetic data. This implies that $T_p$ can also be used to estimate variances of completely synthesised data when this condition holds. When the sample sizes differ for observed and synthetic data the expression for $T_p$ needs modification as
 \begin{equation*}
 T_p = \bar{v}_M k/n+b_M/M.   
 \end{equation*}
Drechsler \cite{Dres_new} has proposed this estimator for ``fully synthetic data", with sampling from the PPD, when both the observed and synthetic data are generated by SRS. 
For completely synthesised data the estimator $T_s$ differs from $T_p$ by replacing $b_M/M$ with $(1+k/n)\bar{v}_M$ for synthesis with PPD and by $(k/n)\bar{v}_M$ without. Since $\bar{v}_M$ will be based on many more degrees of freedom than $b_M/M$ we would expect $T_s$ to be a more precise estimate than $T_p$. This was confirmed by the results of simulations, one of which is reported in Appendix \ref{sec:appendix:srs}. Another advantage of $T_s$ is that it does not require multiple syntheses. 

\subsection{Synthesising data from complex samples}
\label{sec:inference:complex}
In \cite{RRR_2003} the original development of methods for synthetic data assumed that synthetic data would be produced by SRS, whatever the design of the original survey data that were being synthesised. This approach would relieve the analyst with synthetic data of the need to use specialised methods for complex samples for their analyses. We will refer to such methods as complex-to-simple synthesis. In Section \ref{sec:inference:new} we have discussed the case when both the original and synthetic data are generated by SRS, the simple-to-simple case. Agencies providing synthetic data from complex surveys may wish to provide data that can be analysed by the same methods as the observed data: the complex-to-complex case.

Reiter \cite{Reiter_2002} used the complex-to-complex approach in the first paper to evaluated the variance estimator ($T_M$ here, but $T_s$ in \cite{Reiter_2002}). The theory described in Sections \ref{sec:inference:complete} to \ref{sec:inference:incomplete} will apply equally well to samples generated by complex sampling methods, provided the asymptotic assumptions outlined in Section \ref{sec:inference:complete} apply. If the same complex sampling methods are used to analyse the observed and synthetic data then the new simple variance estimators, $T_s$ and $T_{s(PPD)}$, can be used and multiple syntheses will not be required. There are three aspects of the sample design that must be accommodated when analysing a complex sample 1) stratification, 2) weighting and 3) clustering, often in combination. We describe briefly below how these three factors affect the practical task of creating  synthetic data.

\subsubsection{Stratification}
This is the easiest case. The sample is divided into strata and synthesis carried out within each stratum by the methods for SRS, if no weighting or clustering is involved, or by other appropriate methods otherwise. Reiter \cite{Reiter_2002} presents a simulation to evaluate $T_M$ and we extend this in Appendix \ref{sec:appendix:strat} to evaluate the properties of $T_s$ and $T_{s(PPD)}$. Stratifying data before synthesis may also be of benefit, even when it is not required by the design, to preserve important interactions and to avoid computational problems with very large observed data sets.

\subsubsection{Weighting}
Weighting in sample surveys can come about for a number of  reasons. In stratified samples different sampling fractions by stratum require inverse-probability weights. Other design features, such as the selection of only one individual in a household, may also mean that weighting is required to make inferences for the population of individuals. Weighting may also be used to adjust for non-response or post-stratification when the sample is matched to known population totals.

The weights will usually be assumed to be part of $X$, the data that will remain unchanged, although in some cases the weights themselves may be synthesised \cite{mitra_weighting}. In either case, inference will always be conditional on the weights. A   subgroup with weight $w_i$ of size $n_i$  will represent a proportion $w_i/\Sigma{(n_i w_i)}$ of the population and appropriate estimators for design weights are required for inference to the population. To create synthetic data for weighted samples, synthetic samples are generated conditional on the values of the weights. This approach has been used by  \cite{Sakshaug2014} in synthesising data from the American Community Survey adjusting for stratification, weighting and clustering. In general, this method will require that the relationship between the weights and other variables is correctly specified: an aspect of the SDA. Where only stratification is involved then the weights will be constant within strata, so we can be certain that SDA is correct for the weights. If the SDA is met for the data and their relationship to the weights, then analysis by methods that adjust for the design weights for both the observed and synthetic data justifies the use of the new estimators here.

We extend the simulation for stratified sampling in Appendix \ref{sec:appendix:strat} to the case of stratified weighted sampling and demonstrate that $T_s$ and $T_{s(PPD)}$ give unbiased variance estimates in this case.

\subsubsection{Clustering}
Clustering can occur in sample surveys for one of two reasons. It may be introduced as part of the survey design, particularly for door-to-door surveys, as a means of reducing the cost of field work. Alternatively, clustering may be an aspect of the population being surveyed, such as the clustering of characteristics within small areas. 
Surveys with a cluster design are most easily analysed by the ``ultimate-cluster" method, for example as implemented in the \textit{vardpoor} package for \textbf{R}  \cite{vardpoor}. The ultimate-cluster method, which uses cluster-level summary statistics, would not provide a fit that could be used for synthetic data generation. However, the corresponding model-based analyses via hierarchical models can provide such a fit. This method has been used to generate synthetic data \cite{Reiter_2002} and incorporated into a conditional approach by \cite{Sakshaug2014}. These applications have used PPD and the variance estimator $T_M$, but $T_s$ and $T_{s(PPD)}$ could equally well be used if data are to be analysed as a complex sample. 
For the case where interest is in small geographic areas Drechsler and Hu \cite{Dre_Hu_2015} show that using geocoding data can preserve the geographic structure of the data without the need for defining clusters.

\subsection{Summary and recommendations}
The factors which determine which estimators should be used for the variances of estimates from synthetic data are 1) whether the data are completely or incompletely synthesised and 2) whether the methods of inference that would be used for the observed data match those to be used for the synthetic data. 
Importantly, all of the results for inference from synthetic data will only hold if the SDA is true.  Except for simulation studies, the SDA is highly unlikely to be exactly true, so all inference from synthetic data must be considered approximate.
Based on the results above we make the following recommendations for the methods and estimators to use for synthetic data. We consider only the utility of synthetic data, independent of  the impact of methods on the potential for disclosure, which is discussed in Section \ref{sec:disclosure}. 

The decision as to whether the data must be synthesised from the PPD of $Y_{obs}$ depends on whether the methods of inference from the synthetic data are the same as those that would be used to analyse the observed data. Only if this is not the case, is PPD required. Using PPD when it is not required will increase the variance of estimates from the synthetic data. Multiple synthetic data sets $(M>1)$ are necessary to obtain variance estimates when different methods are used for inference from the observed and synthetic data and also for incompletely synthesised data. Increasing M will reduce the variance of estimates from synthetic data, but only to a lower limit of   $U$, the expected variance of $Q$ estimated from the observed data.

For completely synthesised data when the methods of inference for the observed and synthetic data match, the estimator $T_s$ is the best choice. When this is not the case then $T_M$ is the only choice, despite its relatively poor performance. If the original data were generated from a complex sample, and the synthetic data by SRS then information on the design factors, often published in technical reports, may allow $U$ to be approximated from $V$. If this is the case PPD and multiple syntheses would not be required and $T_s$ modified to give
\begin{equation*}
T_{sD_e} = \bar{v}_M (D_e k/n+1/M) 
\end{equation*}
\noindent{where $D_e$ is the approximate design effect. Although this may be inaccurate, it may be more useful in practice than using $T_M$ unless a very large value of $M$ is used.
For incompletely synthesised data the estimator $T_p$ should be used.}

\section{Practical aspects of data synthesis}
 
\label{sec:practical}  
The aim of the methods, as described in the literature we reviewed above, has been to provide synthetic data that can be used for inference to population parameters. In Section \ref{sec:inference} we have also adopted this approach,
but it is not what we envisage for users of the LSs, at least at present. This is also true for users of synthetic data products produced by the U.S. Bureau of the Census, where the publication of results from synthetic data alone is strongly discouraged. 
Staff with access to the original confidential data  create synthetic data which are made available to analysts.  
The analysts carry out data preparation, exploratory analyses and preliminary modelling on the synthetic data. The code developed by the analysts is used to run final fits of candidate models and validation analyses on the original un-synthesised data. Concern about the validity of the SDA underlies these policies.  If this validation proves inadequate, the results can be used to improve the methods to be used for future clients.  The term ``gold-standard" describes the final steps where code for fitting and validating models is run on the actual data.

When this approach is being used the analysts will not be interested in any of the variance estimates discussed in Section \ref{sec:inference}. Instead, they will want an estimate of the results they might get from analysing the observed data. 
This implies that synthetic data must be produced that matches the design of the observed data, i.e. simple-to-simple or complex-to-complex. 
The analysts will want estimates, $\bar{q}_M$ as before, and an estimate of the variance they might get from the observed data which is provided by $\bar{v}_M$. One synthetic data set without sampling from the PPD will be all that is required in this case.
The other estimators discussed here may come into their own if the verification process shows that synthetic data can be relied upon to give results that agree with those from the observed data.

Real survey, census or administrative data may bear very little resemblance to the models used to derive the theory of synthetic data. Continuous data may have distributions that are nothing like a Normal distribution, even after a suitable transformation. Categorical data may have many complex interactions that it would be unreasonable to expect the synthesiser to investigate. 
Furthermore, real data may be subject to constraints that must be respected in the synthetic data. For example, if an analyst were presented with synthetic data on children in families where a natural parent  was less than 12 years older than a child, the utility of the data would be questionable. 
Fortunately these questions have been addressed for synthetic data and the literature contains many options, some of which have been implemented in the $synthpop$ package. Synthesis methods have been developed that preserve the marginal distributions of continuous variables \cite{WandB}  and these can be adapted to include an element of smoothing to prevent the identification of unique values. 
The use of a sequence of conditional distributions makes it easy to incorporate constraints on data values for synthetic samples. Variables that define the constraints must be synthesised first and the constrained variable is then synthesised with the constraint satisfied.

 A number of methods from machine learning have been used successfully to generate synthetic data \cite{reiter_cart, Caiola_and_R_2010, DandR2011}.  Classification and regression tree models (CART) performed well in the evaluation carried out by Drechsler and Reiter \cite{DandR2011}. These have been implemented in $synthpop$ with a choice of algorithms from  the \textbf{R} packages $party$ \cite{party} or $rpart$ \cite{rpart}. We have found that these methods are able to reproduce the main features of data sets without the need for exploratory analysis. One such example is presented in Section \ref{sec:example}. This has led us to set CART as the default method in $synthpop$ if no detailed models are specified by the analyst. Such models may seem distant from the theoretical models used when describing the theory of synthetic data. But the adaptive nature of a series of conditional CART models may give a better approximation to the structure of the observed data than is possible within the constraints of parametric models. They may come under George Box's category \cite{Box} of being ``wrong but useful".

When data to be synthesised have missing values they could be replaced by imputations and the mutiply imputed data sets can then be synthesised. This  has been implemented  \cite{Drechsler_book, SIPP2} using formulae and variance estimates for combining multiple imputations with multiple syntheses  \cite{R_synth_w_imp}. We have not adopted this approach in $synthpop$ because we expect that the choices about handling missing data for a particular project should be the responsibility of the analyst. Using a missing-at-random approach we synthesise the missingness indicator first, and then synthesise the remaining cases from a fit to the non-missing cases in the observed data. Both the synthesised values and the missingness indicator can then be used together in the synthesis of later variables. This guarantees that any relationships with the missingness indicator are  maintained in the synthetic data. For variables earlier in the sequence this is assured by including them as predictors of
 missingness and, for those later in the sequence, by having the missingness indicator as one of their predictors.   An analyst can use synthesised data with missing values to decide how to handle them and their  methods can readily be run on the observed data. In some cases further synthetic data, with missing values ignored or imputed, could be provided to the analyst. 

Data from the LSs often includes time to event data. These may be defined as a series of dates or as a follow-up time and an indicator of  the event at the end of follow up. For example, the LSs are linked to death registrations and emigration records. To synthesise such data the event indicator is synthesised first and the follow-up times are synthesised separately for each type of event. Possible models for follow-up times are parametric survival models (Weibull or log-Normal) or a CART method applied to Kaplan-Meier survival estimates \cite{party}. Cox proportional hazard models would be more difficult to fit because they would require the vector of all observed event times and the corresponding baseline hazard to be used in simulating the synthetic data.
Poisson models can be used for person-years analyses.

\section{Utility and disclosure risk for synthetic data}
\label{sec:disclosure}
Analysis of synthetic data will never yield exactly the same results as would be found from the observed data. 
We hope synthetic data will be useful and  the term ``utility" is used for the extent to which results from synthetic data agree with those from the original data. 
Utility measures may be specific to the inferences being made from the data or they may be general utility measures that assess the overall difference between the observed and synthesised data distributions. In both cases the utility of synthetic data will depend crucially on the SDA. Analysis-specific utility measures, such as confidence interval overlap, are appropriate for evaluating the results of a final analysis as is done with a verification server \cite{verserv,MclRVer}. General utility measures can provide the person creating the synthetic data with a means of comparing utility for different ways of synthesising data before any analyses have been carried out.   Proposals for general utility measures have been suggested by Woo et al. \cite{WooProp} and Snoke et al. \cite{snoke} have adapted them  for the specific case of synthetic data. Functions to compute these utility measures are being incorporated into the $synthpop$ package.

Disclosure risks measures have been developed for incompletely synthesised data  \cite{Reit_mit,Dre_Rei} that use the information in the unchanged variables to identify links to known observations in the observed data. 
Although completely synthesised data contain no records that are identified  with real cases, this does not mean that their real or perceived disclosure risk is zero. 
The development of $\epsilon$-differential privacy \cite{dwork} provides a measure of disclosure risk which does not depend on the specific tactics of an intruder, although it assumes the intruder possesses a very large amount of ancillary knowledge about the real data. The method of creating synthetic data and making inferences from them, as described here, is not guaranteed to satisfy differential privacy and attempts to modify it for simple examples \cite{Char, A_V, McCl_Rei}  can severely damage its utility.  An extension to this method, $\epsilon$-$\delta$-differential privacy \cite{ono, dwork_smith}, which allows a mechanism (e.g. a synthesis) to satisfy $\epsilon$-differential privacy with probability $1-\delta$, has been used to generate the synthetic data that sit behind the U.S. Bureau of the Census's ``OnTheMap" application\footnote{U.S. Census Bureau. 2015. OnTheMap Application. Longitudinal-Employer Household Dynamics Program. \url{http://onthemap.ces.census.gov/}}. But this also damaged utility and the results from the synthesising algorithm required extensive adjustment \cite{otm}. 

Elliot \cite{Elliot_2014} has evaluated output from the $synthpop$ package using data from the UK Living Costs and Food survey  by a related method termed ``empirical differential privacy" but the methods are specific to the type of attack that an intruder is presumed to be attempting. 
More recent research with synthesised categorical data \cite{Hu_etal,ReiWangZhang,Mcclure_reiter_2015} has proposed other methods that can be used to identify individual records with a high disclosure potential, but these methods cannot at present provide measures that can be used with sort of complex data that we are synthesising.

How will our recommendations for the analysis of completely synthesised data affect any disclosure risk posed by  the synthetic data? By not sampling from the PPD we are introducing less noise than would be the case with PPD and thus might expect to increase the disclosure risk. But the need to generate only a single synthetic data set will reduce the disclosure risk \cite{Reit_mit,KleinSinha2015c}.  Our proposals to supply synthetic data to users had to be agreed by the SLS Research Board and we have not asked for permission to release more than one synthetic data set to a user. Synthetic data from the SLS will be released to accredited and trained users who have signed a disclaimer that they will not identify individuals, nor share the data with anyone not accredited for their specific project. This will reduce the possibility that disclosure harm will be associated with any disclosure potential \cite{CS}.

The report mentioned above \cite{Elliot_2014} formed part of our submission to the SLS Research Board, but another  aspect of the release of synthetic data was of concern to them. If an intruder sees the synthetic data they may mistakenly believe it to be real and attempt an identification, with subsequent loss of reputation for the data collection agencies, even when no actual disclosure has taken place. We have agreed to carry out additional  measures to avoid any such problem. These  are now incorporated into the $synthpop$ package. All data sets have to be clearly labelled as ``false data". Any observations that are unique in the actual data and also appear as unique in the synthetic data have to be removed from the latter. Options for smoothing of continuous data and top-coding of outlying observations are also available in $synthpop$.
 
To evaluate synthetic data thoroughly we need quantitative measures of both disclosure risk (R) and data utility(U) that would allow different synthesis methods to be compared on an R-U confidentiality map (\cite{DandF1999} and \cite{DuncEllSal}, Chapter 6).  Such measures have been developed for other disclosure control methods \cite{DFetal2001,trotF2002,kim2016,DFT} and permit evaluation of the R-U trade-off.  In the $synthpop$ package we have implemented measures of utility (U), but a measure of disclosure risk (R) is not currently available for data with the complexity of most survey and administrative data. We hope that further research in this area will help us to meet this challenge in future.

\section{Example}
\label{sec:example}
\subsection{Methods}
\label{sec:methods}
The fitting of formal models is only a small part of any statistical analysis. The majority of an analyst's time is taking up with checking and exploring the data and in carrying out preliminary tabulations. We aim to produce synthetic data that can be used for this type of analysis. 
To illustrate synthesis of LS data we have extracted data on age, sex, marital status, ethnic group and long-term illness from the SLS database for the 1991 and 2001 Censuses. The acronyms AGE9, SEX9, MSTAT9, ETH9, ILL9 and AGE0, SEX0, MSTAT0, ETH0, ILL0 are used to describe them.  The synthesis was carried out for over 186 thousand SLS members who were present at both censuses. No preliminary data cleaning was carried out on the extract. Some variables had a small percentage of missing values. For the categorical variables the missing cases are simply handled as an additional category, but AGE0 had 0.15\% of missing cases coded as -999  and this formed a missingness indicator for AGE0.  Univariate distributions, cross tabulations and results from fitted models were compared for the observed and the synthetic data. The $synthpop$ package was used for all analyses and more details of the methods can be found in \cite{synthpop_jss}. 

Initial synthesis was carried out without PPD for two choices of models. The first (CART) used CART for all the variables and the second (Parametric) used an appropriate parametric method
for each variable: polychotomous or logistic regression for categorical data and distribution-preserving
linear regression for AGE9 and AGE0. A lower triangular prediction matrix was used in both cases
so that all previously synthesised variables were used in the predictions. The ordering of the variables
used in the final syntheses and for the results presented here was ETH9, ETH0, AGE9, SEX9, MSTAT9, ILL9, AGE0, SEX0, MSTAT0, ILL0. The two ethnic group variables each had a fairly large number of categories (35 and 16). Initially they were placed towards the end of the synthesis order, but this slowed down the parametric synthesis because the multinomial models required a very large number of iterations for convergence. Moving them to the start of the synthesis overcame this.

\subsection{Results of exploratory analyses} 
\begin{figure}[h]
	\captionsetup{width=.9\textwidth}
	\centering
	\vspace{1cm}
	\includegraphics[height=2in]{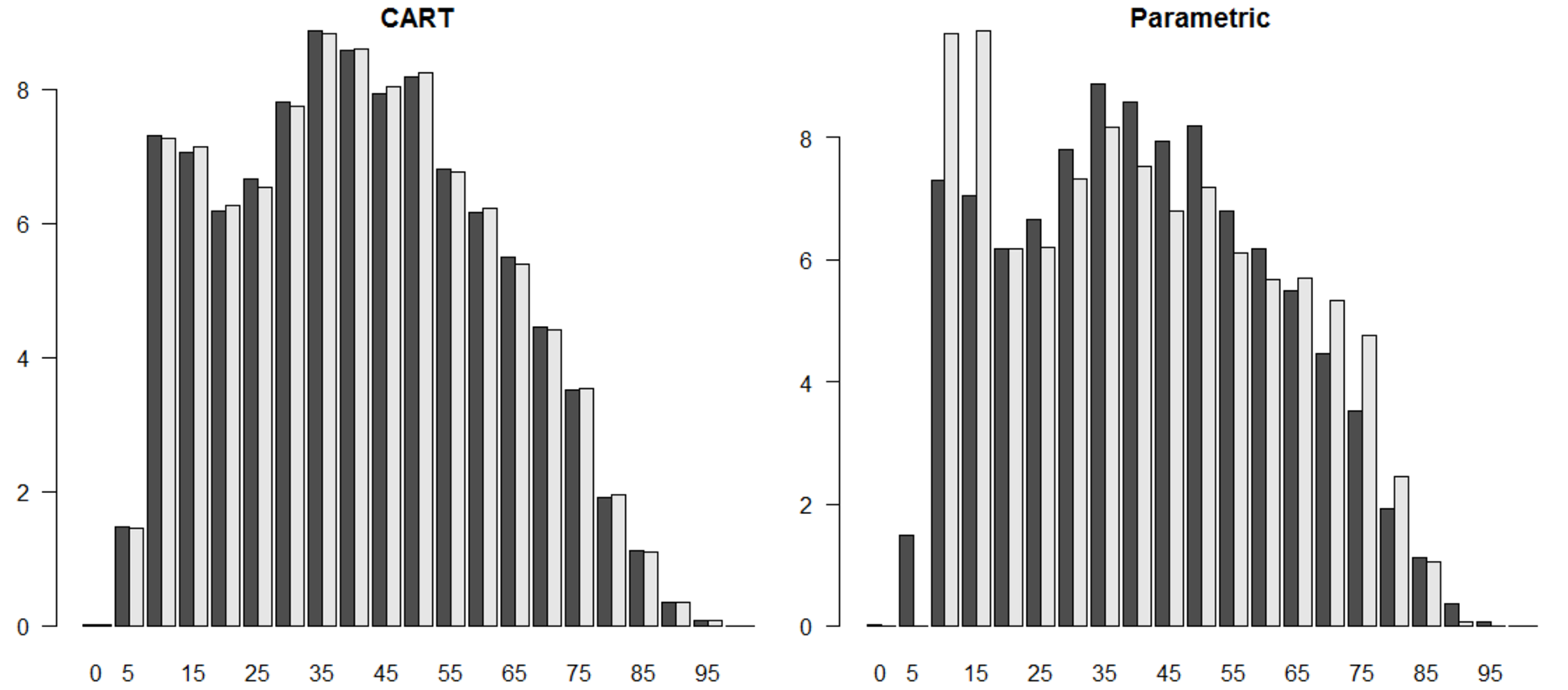}
	\caption{\label{fig01}Comparison of observed (black) and synthetic (grey) data for AGE0 in 5 year age groups.}
\end{figure}

\begin{figure}[h]
    \captionsetup{width=.9\textwidth}
	\centering
	\makebox{\includegraphics[height=2in]{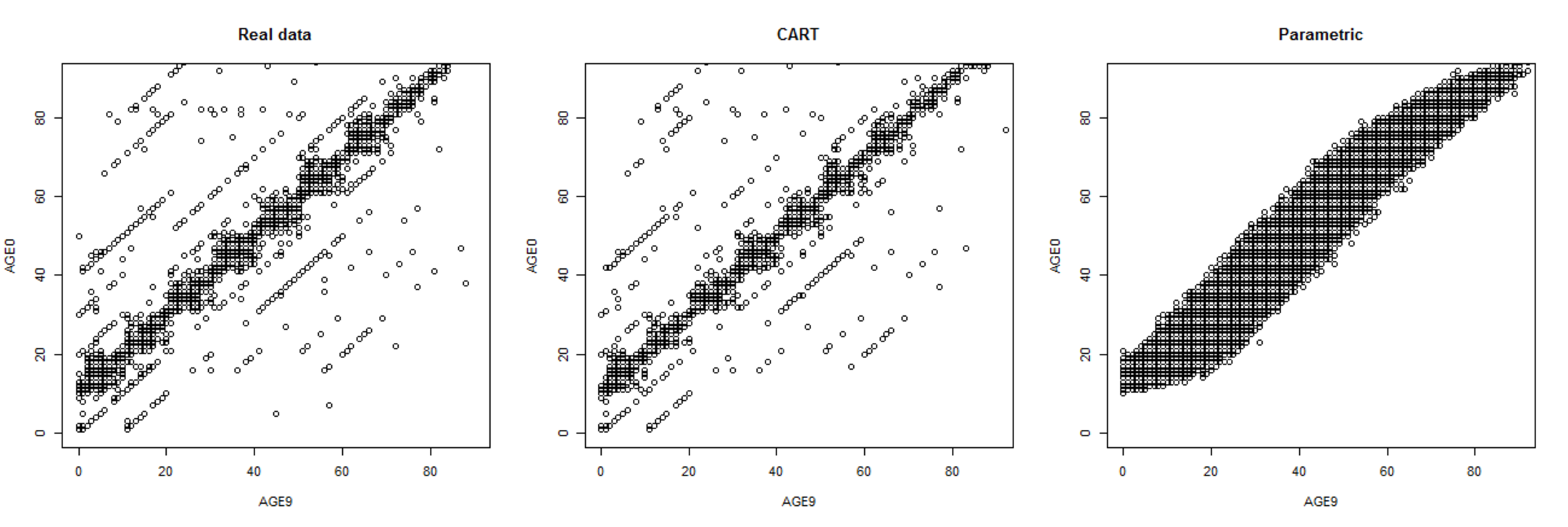}}
	\caption{\label{fig02}Plots of AGE9 against AGE0 for observed data and one synthesis by each of CART and parametric methods (most points refer to multiple observations).}

  \begin{minipage}{6.3in}%
    \vspace{0.1cm}
    \small \textit{Notes}: most points refer to multiple observations; points at ages over 90 are not shown to avoid any possible disclosure of extreme ages.%
    \vspace{0.2cm}
  \end{minipage}%

\end{figure}

A problem with the initial run of the synthetic data was that some SLS members under 16 had marital status ``married" in the synthetic data, with the number of such cases being larger for the Parametric syntheses. This was readily fixed by imposing a logical rule during the syntheses. 
Marginal distributions of all the variables were comparable to those for the observed data for syntheses by CART, but for parametric methods the fit to the marginal distributions of AGE0 was unsatisfactory (Fig. \ref{fig01}).

 From Figure \ref{fig02} we can see why the parametric method failed to reproduce the distribution of AGE0. Age is recorded in full years and, in most cases, AGE0 is exactly AGE9+10. Exceptions could be those with birthdays between the dates of the two censuses as well as various data errors or mismatches. Differences of exactly 10 or 20 years are common. The parametric method could not reproduce this pattern. The parametric syntheses were rerun with the method for just this variable changed to CART giving satisfactory results. 
Thus for these exploratory analyses the CART method gives  more satisfactory results than parametric methods, and with no requirement to customise the analyses in any way.

\subsection{Results of fitting models to data syntheised without PPD}

\begin{figure}[h]
	\captionsetup{width=.9\textwidth}
	\centering
	\vspace{1cm}
	\makebox{\includegraphics[height=3in]{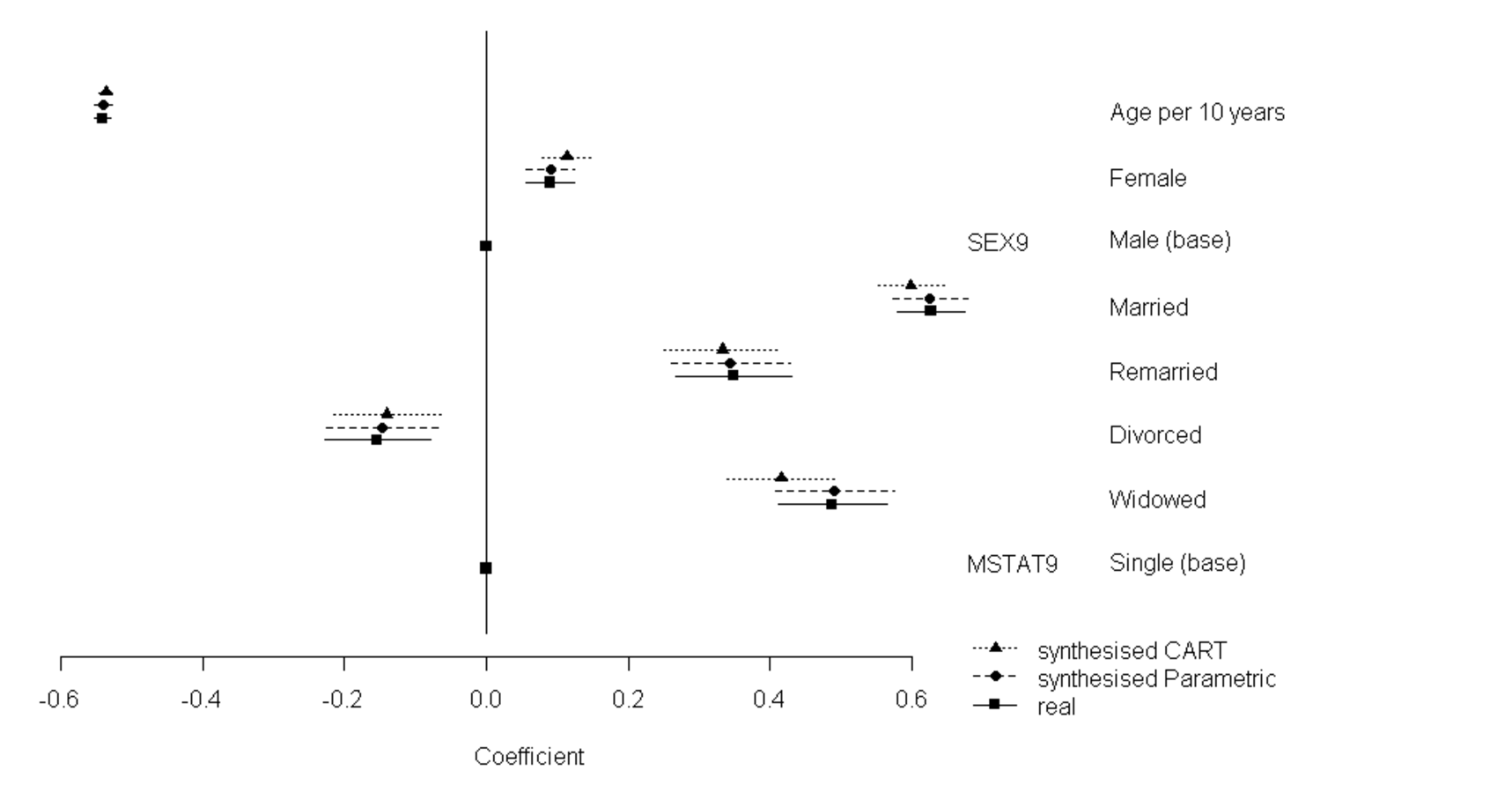}}
	\caption{\label{fig03}Coefficients of fit to ILL9=``No" from AGE9, SEX9 and MSTAT9 for observed and synthetic data.}
\end{figure}
We now investigate how inference from synthetic data differs from that from the observed data using variables from the data set described in the last section. 
A logistic regression modelled absence of long-term illness in 1991 (ILL9) from AGE9, MSTAT9 and SEX9.  
We use the same two synthesis models (Parametric and CART) described above with $M=10$ in each case, the rule for marital status for the under 16s in place and ILL9 as the last variable synthesised. 
Initially a model with the main effects of each variable was fitted to the observed and the synthetic data sets.
Figure \ref{fig03} compares the estimates from the observed data with the averages from the 10 simple syntheses from parametric and CART  models. Confidence intervals from the synthetic data sets are calculated from  $\bar{v}_M$, the estimate of the variance that would be expected from the observed data. Since the final step in the synthesis was a logistic regression of ILL9 on these three variables, we know that the SDA will be met for this model. As expected, the estimates from the parametric model are very close to the observed data estimates, and a formal test showed that there was no evidence of any bias. In this case the estimates from the CART syntheses are also fairly close and certainly would be satisfactory for exploratory analyses.  We can see that freedom from long-term illness decreases sharply with age and is higher for females than males. Adjusting for age and sex, those married, remarried or widowed are more likely to be free from long-term illness than those who are single, whereas the opposite is true for the divorced.

A further model was fitted which includes a sex by marital status interaction, so that the SDA will not hold for the synthesis from our Parametric models. 
Results are shown in Figure 4. For the observed data there is evidence of an interaction. 
The association of being married with lack of illness is stronger for men than for women. The CART syntheses do a reasonable job of reproducing this, whereas the parametric syntheses show no evidence of this interaction since they are generated from an interaction-free model. The analyst with access only to the synthetic data generated by this parametric model would  find no evidence of this interaction in any diagnostic plots. The CART model may be less exact when the parametric model is correct, but its flexibility shows evidence of an interaction from the synthetic data.

\begin{figure}[h]
	\captionsetup{width=.9\textwidth}
	\centering
	\makebox{\includegraphics[height=3.2in]{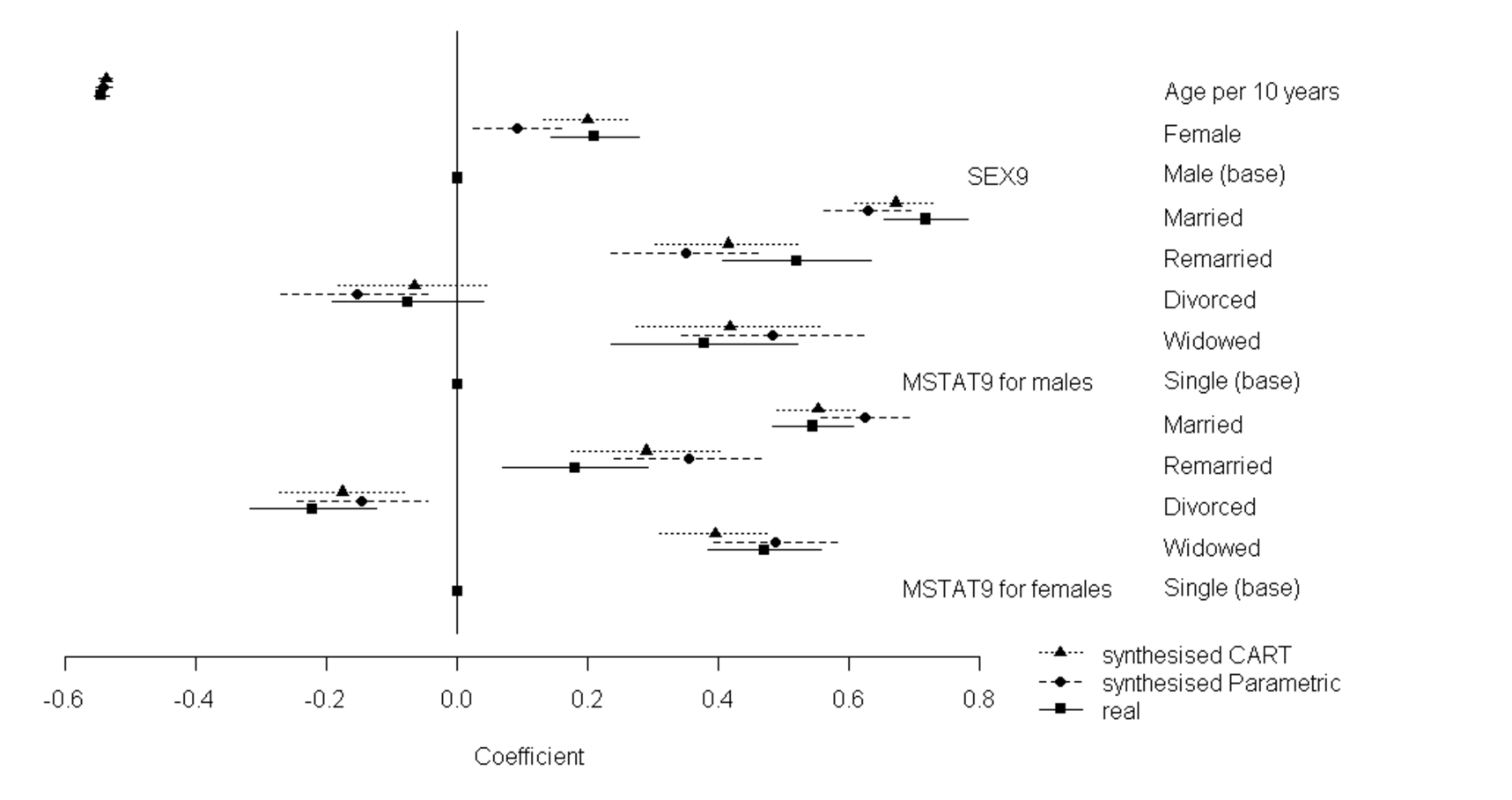}}
	\caption{\label{fig04}Coefficients of fit to ILL9=``No" from AGE9, SEX9 and  MSTAT9*SEX9 interaction for observed and synthetic data.}
\end{figure}

These results all assume that the analyst is interested in estimating the results which might be obtained from the observed data, rather than in making inferences to population parameters directly from the synthetic data. We believe this is an appropriate use of synthetic data because we can never be completely sure that the SDA is met. Further experience with synthetic data may show that this is too cautious a position, but it is a safe one for now.  

\subsection{Inference to population parameters from data generated with and without PPD}

Finally,  we present inferences for the population parameters  for this example to illustrate the properties of the variance estimates described above. We  fitted the same model as in the previous section, with just the main effects, with parametric and CART methods and compared variance estimates with those from the observed data. CART synthesis with PPD starts by taking a bootstrap sample of the observed data and fits the CART models to this sample. For synthesis without PPD we have a single  estimate  $T_{s}$ whereas, for synthesis with PPD, we have three variance estimates $T_{s(PPD)}$, $T_{p}$ and $T_{M}$. Syntheses with parametric and CART models were carried out with $M=10$ in each case. 

The synthetic estimates of the coefficients all showed the same patterns for CART and parametric methods as in Figure 3.
The expected values of the variances of $\bar{q}_{M}$ as an estimate of $Q$ should be $V(1+1/M)$ without PPD and $V(1+2/M)$ with PPD, since $k=n$. Table \ref{tab02} gives the ratios of the standard errors divided by $\sqrt{V}$, estimated from the standard error from the observed data.  If the SDA holds the expected value of these ratios of standard errors would be approximately 1.049 without PPD and 1.095 with PPD. 
For the CART syntheses, we can see that standard errors calculated from $T_{s}$ and $T_{s(PPD)}$ are close to their expectations. Those calculated from $T_{p}$ are somewhat larger for the CART methods. The results are more irregular for the parametric methods which gave more biased estimates of the standard errors. In both cases, as expected, the estimator from $T_{M}$ is quite unsatisfactory, giving a negative value in one case. 
These results support our recommendation of the use of the variance estimates $T_{s}$  for completely synthesised data.

\begin{table}[h]
	\captionsetup{width=.9\textwidth}
	\caption{\label{tab02} Ratios of standard errors calculated from different variance estimates to the standard errors from the observed data.  Inference for population coefficients of a logistic regression of  ILLP9 predicted from AGE9, SEX9 and MSTAT9 from synthetic data produced by parametric and CART methods, with and without PPD.}
	\centering
	
	\begin{tabular}{{l}*{8}{c}}
		\hline
		\noalign{\vskip 3px}
			\noalign{\vskip 3px}
		& \multicolumn{4}{c}{Parametric} & \multicolumn{4}{c}{CART}\\
		& Without PPD & \multicolumn{3}{c}{With PPD} & Without PPD & \multicolumn{3}{c}{With PPD}  \\
		& $\sqrt{T_{s}}$ & $\sqrt{T_{s(PPD)}}$ & $\sqrt{T_{p}}$ & $\sqrt{T_{M}}$ & 	 $\sqrt{T_{s}}$ & $\sqrt{T_{s(PPD)}}$ & $\sqrt{T_{p}}$ & $\sqrt{T_{M}}$ \\  
		\noalign{\vskip 3px}
		\hline
		\noalign{\vskip 3px}
		Intercept & 1.046 & 1.091 & 1.078 & 0.940 & 1.045 & 1.098 & 1.107 & 1.201 \\ 
		AGE9 & 1.165 & 1.217 & 1.196 & 0.952 & 1.048 & 1.095 & 1.133 & 1.458 \\  
		SEX9 & & & & & & & &  \\
		Female & 1.043 & 1.089 & 1.174 & 1.820 & 1.049 & 1.095 & 1.131 & 1.447 \\ 
		MSTAT9 &  & & & & & & &   \\
		Married  & 1.172 & 1.225 & 1.248 & 1.465 & 1.048 & 1.097 & 1.232 & 2.159 \\ 
		Remarried & 1.086 & 1.135 & 1.103 & 0.704 & 1.052 & 1.095 & 1.119 & 1.337 \\ 
		Divorced & 1.091 & 1.143 & 1.081 & NA & 1.064 & 1.089 & 1.212 & 2.076 \\ 
		Widowed & 1.145 & 1.199 & 1.208 & 1.294 & 1.048 & 1.099 & 1.216 & 2.044 \\ 
		\noalign{\vskip 3px}
		\hline
		\noalign{\vskip 3px}
		\multicolumn{9}{l}{ $ Note $: NA -- Negative variance estimate}\\
	\end{tabular}
\end{table}

\section{Summary and future directions}

It is now over twenty years since the first proposals were made to use synthetic data for disclosure control. A rich literature has since been developed, mainly by Reiter and his colleagues on the theory and practice of synthetic data generation. We were able to draw on this in developing the $synthpop$ package. While we make some suggestions for different interpretations of the methods and introduce some new variance estimators, our work depends heavily on their insights.

There is much more we still need to learn about the best way to carry out syntheses. Recommendations are needed on choosing the ordering of variables during synthesis, for deciding whether reduced models excluding some variables should be used and for fine-tuning the parameters of CART models.  The $synthpop$ package is intended to facilitate this and new methods can be added by the user that are not at present part of the package. 
The synthesiser, with access to the observed data, ought to carry out checks on the validity of the data before it is released to the analyst.  At a minimum, a visual check on all the marginal distributions should be carried out. The $synthpop$ package includes a function to do this and Figure \ref{fig03} is an example of part of its output. Now that registered and accredited users of the SLS are being supplied with synthetic data, their feedback on its performance will  help us to develop best practice. We encourage users to include code to  verify any assumptions (such as the absence of interactions) when a gold-standard analysis is run.

The structure of administrative data also presents challenges. In particular, we hope to develop methods for repeat events, such as hospital admissions and for the synthesis of data for all members of a family. Options to include utility measures are already being provided in $synthpop$, but the ability to include measures of disclosure risk remains a more distant goal.

\appendix
\section{Appendix} \label{sec:appendix}

\subsection[app1]{Simulation  to evaluate methods for synthetic data with simple random sampling}
\label{sec:appendix:srs}
This simulation is similar to one used to evaluate methods for synthetic data and the variance estimator
$T_M$ in  \cite{RRR_2003}. The population was created by drawing a sample of size $ N = 50,000 $ from  $N(0,\Sigma)$, where
$\Sigma$ is a 5 by 5 matrix with diagonal elements 1 and off diagonals 0.5. The columns of the population
are denoted by $(y_{1},y_{2},y_{3},y_{4},y_{5})$. Each of 10,000 simulations selected a random sample of size $ n = 500 $
from this population as the observed data. The parameters of the five-variate Normal distribution were estimated for each simulation and two sets of synthetic data were generated, each with $ M = 5 $ replicates
of size $ k = 1,000 $, one set with PPD and one without. For the observed data the fpc is close to unity here,
at 0.99. Simulations where this was not the case gave estimated variances that were biased, illustrating
the need for this assumption to be met, as discussed in Section \ref{sec:inference:complete}.

A linear model for $y_{1}$ predicted from the other 4 variables
was used to evaluate methods of
inference. The simulations without PPD confirmed that
$T_s$ and $T_p$ are unbiased estimates of the variance of the coefficients. The same was true for $T_M$, $T_{s(PPD)}$ and $T_p$
from the simulations with PPD. But $T_M$ was only unbiased because of the inclusion of approximately
11\% of results with negative variances, which would never be accepted in practice. Two approaches were used to adjust the negative values of $T_M$;
basing the intervals on only the cases where $T_M > 0$ or using an an estimator $T_M^{adj}$, proposed by Reiter \cite{Reiter_2002}. In both these cases the variance estimates were biased upwards by 13\% and 10\% respectively.

\begin{table}[ht]
    \captionsetup{width=.9\textwidth}
	\centering
	\caption{\label{tab02} Ratio of the variance from the simulations for existing estimators compared to the variance of the new estimators $T_s$ and $T_{s(PPD)}$.}
		\begin{tabular}{lccccc}	
			\hline
			\noalign{\vskip 3px}  
			& \multicolumn{1}{c}{Without PPD} & \multicolumn{4}{c}{With PPD}\\
			& \multicolumn{1}{c}{Ratio to variance of $T_s$} & \multicolumn{4}{c}{Ratio to variance of $T_{s(PPD)}$  }\\
			& $T_{p}$ & $T_p$  & $T_M$ & $T_{M}^{adj}$ & $T_{M}>0$ \\
			\noalign{\vskip 3px}  
			\hline
			\noalign{\vskip 3px}  
			Intercept & 1.94 & 5.68 & 173.44 & 145.47& 161.97\\
			y2        & 1.49 & 3.54 & 94.77  & 81.13 & 89.19\\
			y3        & 1.45 & 3.52 & 94.27 &  80.25 & 88.61 \\
			y4        & 1.45 & 3.52 & 94.14 &  86.30 & 85.80 \\
			y5        & 1.48 & 3.49 & 93.65 &  80.15 & 88.01\\
			\hline
		\end{tabular}
	\end{table}

The variance, estimated from the simulations, of $T_M$ (and its adjusted variants) and Tp, relative
to the new simple estimates are shown in Table 2. The very poor precision of all the $T_M$ estimators is
evident. The new simple estimators $T_s$ and $T_{s(PPD)}$ have better precision than $T_p$. Table 3 shows the estimated coverage calculated for all these variance estimates, except for $T_M$,
where the negative values prevent the estimate of coverage. In practice, an analyst faced
with a negative variance estimate, would either abandon the whole enterprise and perhaps ask for a
new set of synthetic data, or use the adjusted estimator $T_M^{adj}$. We can see that all the results are very
satisfactory except for the adjusted $T_M$ estimators, which have reduced coverage.
 
\begin{table}[ht]
	\captionsetup{width=.9\textwidth}
	\centering
	\caption{\label{tab03} Coverage of 95\% confidence intervals for $Q$ calculated from different variance estimators.}
	\begin{tabular}{rrrrrrrr}  
		\hline
		\noalign{\vskip 3px}  
		& \multicolumn{2}{c}{Without PPD} & & \multicolumn{4}{c}{With PPD}\\
		& $T_{s}$ & $T_{p}$ & & $T_{s(PPD)}$ & $T_{p}$ & $T_{M}^{adj}$ & $T_{M}>0$ \\
		\noalign{\vskip 3px}   
		\hline
		\noalign{\vskip 3px}  
		Intercept & 94.7 & 94.7 & & 95.0 & 94.7 & 86.1 & 85.4 \\ 
		y2 & 95.4 & 95.3 & & 95.0 & 94.8 & 96.8 & 86.3 \\ 
		y3 & 95.1 & 95.2 & & 95.2 & 94.9 & 86.5 & 85.7 \\ 
		y4 & 94.9 & 94.9 & & 94.8 & 94.4 & 86.3 & 85.8 \\ 
		y5 & 95.3 & 95.2 & & 95.0 & 94.7 & 86.6 & 86.1 \\ 
		
		\hline
	\end{tabular}
\end{table}

\subsection{Simulation to evaluate synthesis for a stratified sample}
\label{sec:appendix:strat}

\subsubsection{Methods}
This simulation was based on that presented in Section 3.2 of \cite{Reiter_2002} where
the simulated observed data and the synthetic data are drawn as stratified samples. A population of size $ N = 10,000 $ was created consisting of 10 strata $h = 1,2,...,10$ each of size $N_h =1,000$
and where the variable $ y $ of interest is distributed as $N(10h, h^2)$, in the $h^{th}$ stratum.
Each simulation selected a stratified random sample of total size $n = 200$, with $n_h$ from each stratum, from this population  to represent the observed data.  Synthetic samples of size $200$ were then generated. In \cite{Reiter_2002} this was done by first taking a
random sample from the population without constraining it to balance the population totals. 
In the simulation reported here exactly $n_h = 20$ values were synthesised for each stratum so that inference from the synthetic data will be the same as from the observed data. The information on stratum membership, which is fixed for each synthetic sample is the $X_{obs}$ for this example, and all inference is conditional on $X_{obs}$.

The parameter ($Q$) to be estimated is the population mean  estimated by the stratified estimate $\sum_{h=1}^{10}{(N_h/N) \bar{y}_h}$ where $\bar{y}_h$ is the mean for of the observations in the $h^{th}$ stratum with variance estimated from (1). 
\begin{equation*}
\sum_{h=1}^{10}{ {\left(1-\frac{n_{h}}{N_h}\right)\left(\frac{N_h}{N}\right)^2}}
\frac{\sum_{j=1}^{n_h}{\left(y_{hj}-\bar{y}_{h}\right)^2}}  {\left(n_{h}-1\right)n_{h}}
\end{equation*}

In \cite{Reiter_2002} the value of $n_h$ was 20 for all $h$, so no weighting is involved, and $ M = 100 $ synthetic data sets were produced for each simulation. This sample design has an extreme design effect. If the usual SRS formula for the variance of the mean were to be used the result would have a variance more that 22 times greater than if the correct one were used. We have extended the simulation to a more realistic number of syntheses ($M=10$) and to different sampling fractions in each stratum. For each of the following three cases we ran 1,000 simulations of synthesis with and without PPD:
\begin{enumerate}
\item{ Unweighed simulation with $M=100$ and all $n_h  = 20$,  as in \cite{Reiter_2002}.}
	\item{As the first simulation but with $M=10$.}
	\item{Selecting samples of size  11, 13, 15, 17, 19, 21, 23, 25, 27, 29 from strata 1 to 10. This gives weights that range from 91 to 34. The larger sampling fraction in the more variable strata should give a reduced variance, although to a lesser extent than the optimal sampling scheme which would give much more extreme weights.}
\end{enumerate}

\subsubsection{Results}

All the simulations gave unbiased estimates of $Q$. For synthesis with PPD simulations 2 and 3 had 6.5\% and 6.2\% of results with negative estimates for $T_M$, while there were no negative estimates for simulation 1 with $M=100$. For simulations 2 and 3, only the adjusted estimates from $T_M$ could be used for confidence intervals.

Table \ref{tab04} compares the variances from the simulations with the mean of the different estimators as well as the coverage  of nominal 95\% intervals for based calculated from each estimator. The variance estimate $T_M$ from proper synthesis  is slightly biased upwards, as was found in the original publication. This may relate to the use of the fpc which would be valid for the observed data, but not appropriate for the synthetic data. All the other estimators appear unbiased, at least to the accuracy provided by the simulation. The adjustments to $T_M$ required in simulations 2 and 3 resulted in increased variances and lower confidence interval coverage. The coverage of intervals based on all other variance estimators appears satisfactory.

\begin{table}[ht]
	\captionsetup{width=.9\textwidth}
	\centering
	\caption{\label{tab04} Variance of the estimates of the mean from stratified sampling from the simulations compared to the average of the different variance estimators and the coverage of the corresponding of 95\% confidence intervals.}
	\begin{tabular}{lrlrlrl}
		\hline
		\noalign{\vskip 3px}  
		& \multicolumn{2}{c}{Simulation 1} & \multicolumn{2}{c}{Simulation 2} &
		\multicolumn{2}{c}{Simulation 3} \\
		& Variance & Coverage & Variance & Coverage & Variance & Coverage \\[.3ex] 
		\hline 
		\noalign{\vskip 3px}   
		Without PPD \\
		From simulation & 0.200 &  & 0.220 &  & 0.166 &  \\ 
		$T_s$ & 0.189 & 94.1 & 0.206 & 94.9 & 0.173 & 94.5 \\ 
		$T_p$ & 0.189 & 94.1 & 0.206 & 94.5 & 0.174 & 94.5 \\[1.5ex]
		
		With PPD\\
		From simulation & 0.197 &  & 0.242 &  & 0.191 &  \\ 
		$T_{s(PPD)}$ & 0.191 & 94.6 & 0.225 & 93.9 & 0.189 & 94.5 \\ 
		$T_p$ & 0.191 & 94.7 & 0.229 & 93.8 & 0.193 & 94.8 \\ 
		$T_M$ & 0.244 & 95.7  & 0.279 &  & 0.237 &  \\ 
		$T_M^{alt}$ & 0.244 & 95.7 & 0.294 & 88.9 & 0.249 & 90.1 \\ 
		$T_M>0$ & 0.244 & 95.7 & 0.302 & 88.8 & 0.256 & 90.1 \\
		\noalign{\vskip 3px} 
		\hline
	\end{tabular}
\end{table}

\bibliographystyle{acm}
\bibliography{applied_stats}
   
\end{document}